\documentclass[aps,prb,preprint]{revtex4-1}
\usepackage{amsmath,amssymb,epsfig}

\begin{document}

\title{Pentagon chain in external fields}

\author{Gy\"orgy~Kov\'acs$^{}$, 
and Zsolt~Gul\'acsi$^{}$}
\address{Department of Theoretical Physics, University of Debrecen, 
H-4010 Debrecen, Hungary}

\date{August 7, 2015}

\begin{abstract}
We consider a pentagon chain described by a Hubbard type of model considered
under periodic boundary conditions. The
system i) is placed in an external magnetic field perpendicular to the
plane of the cells, and ii) is in a site selective manner under the action of 
an external electric potential. In these conditions we show in an exact 
manner that the physical properties of the system can be
qualitatively changed. The changes cause first strong modifications of the
band structure of the system created by the one-particle part of the 
Hamiltonian, and second, produce marked changes of the phase diagram. We
exemplify this by deducing ferromagnetic ground states in the presence of
external fields in two different domains of the parameter space. 
\end{abstract}
\pacs{71.10.Fd, 71.27.+a, 03.65.Aa} 
\maketitle


\section{Introduction}

Pentagon chains as main representatives of conjugated polymers which have 
conducting properties have 
broad application possibilities covering several fields. Starting from the 
Nobel prize awarded to the subject in 2000, in 15 years several
technological applications have emerged, such as
energy storage \cite{App7}, solar cells \cite{App9}, or sustained drug 
release \cite{App11}, which are now considered classical.
But besides these, several other application possibilities are present
as rechargeable batteries, electrochromic displays, information 
memory, anti-static materials, anti-corrosives, electrocatalysis, sensors, 
electromechanical devices, infra-red polarizers \cite{App0},
bioanalytical sensors \cite{App1}, or  biocontacts \cite{App3}.

In addition to these, there are several other
applications related explicitly to the action of external (magnetic 
and/or electric) fields on the conducting polymer of pentagon chain type. 
In order to exemplify, we mention the applications related to
pseudocapacitor electrodes \cite{App2},
fabrication of cell layers creating in vivo electric 
fields \cite{App4}, radar or microwave absorbing materials which converts 
the electromagnetic energy into heat \cite{App5}, electrods for battery 
applications \cite{App6}, artificial muscles based on actuators with 
large strain and stress induced electrically \cite{App8},
photothermal agents \cite{App10}, covering layers for magnetic 
nanoparticles \cite{App12}, or transport mediators in liquid electrolites 
\cite{App13}.

Contrary to this technological interest, the action of external electric
and/or magnetic fields on pentagon chains has not been analyzed et all on 
microscopic level. The present paper tries to fill up at least partially
this gap by reporting a detailed study of the action of external magnetic 
and electric fields on pentagon chains at microscopic level. 

We mention that pentagon chains in the absence of external fields have been
intensively analyzed in the last period. Using technical steps presented in
details in Ref. \cite{Intr8}, the high concentration limit has been described
\cite{Intr25,Intr25a}, and the low concentration limit has been 
investigated \cite{MT3,MT3a,MT3b,MT3c}. Inspite of the fact that these systems 
are non-integrable, and strongly correlated \cite{STR1} where the on-site 
Coulomb repulsion may even be as large as 10 eV \cite{STR2}, the method used 
is special and provides exact
results \cite{Intr8,STR3}. The strategy is to transcribe the Hamiltonian in 
exact terms into a
positive semidefinite form $\hat H = \hat O +C$ where $\hat O$ is a positive
semidefinite operator while $C$ is a scalar, and in a second step to find
the exact ground state by constructing the most general solution of the equation
$\hat O |\Psi\rangle =0$. By deducing total particle number ($N$) dependent 
ground states in this manner,
we can find exact result relating the low lying part of the excitation spectrum
as well, e.g. via the N-dependent chemical potential.
The technique itself has been successfully used previously in characterizing
different non-integrable quantum many-body systems as periodic Anderson model 
(PAM) in one \cite{PAM1a,PAM1b}, two \cite{PAM2a,PAM2b}, and three
\cite{PAM3a,PAM3b} dimensions, non-integrable chain structures \cite{Q1},
two dimensional effects as stripe or droplet formation \cite{D1},
delocalization effect of the interaction \cite{D2}, or even study of
the effect of randomness \cite{D3}.

Using the above presented procedure, in the present paper we show that 
placing the pentagon chain in external fields, main physical properties of
the system can be substantially changed.

The remaining part of the paper is structured as follows: Section II.
describes the studied systems in external fields, Section III. presents the
system Hamiltonian, Sections IV-V. present the deduced ground states
in two different regions of the parameter space,
Section VI. containing the summary closes the presentation, while at the end of 
the paper two Appendices containing mathematical details close the presentation.

\section{The system under consideration placed in external
magnetic and electric fields}

The system under consideration is a pentagon chain taken with periodic
boundary conditions and containing $N_c$ unit cells as presented in Fig.1.

\subsection{The effect of the magnetic field}

We consider the chain in an external magnetic field perpendicular to the
plain containing the cells. In these conditions each hopping matrix element 
$t_{i,j}=t_{i \to j}$
will gain a Peierls phase factor $e^{i\phi_{i,j}}$. This means that along the bond
$(i,j)$ described at $B=0$ by $t_{i,j}$, the hopping matrix element for 
$B \ne 0$ case becomes $e^{\phi_{i,j}}t_{i,j}$. The Peierls phase factor is given 
by an integral along the bond $(i,j)$ of the form

\begin{figure} [h]                                                         
\centerline{\includegraphics[width=15cm,height=10cm]{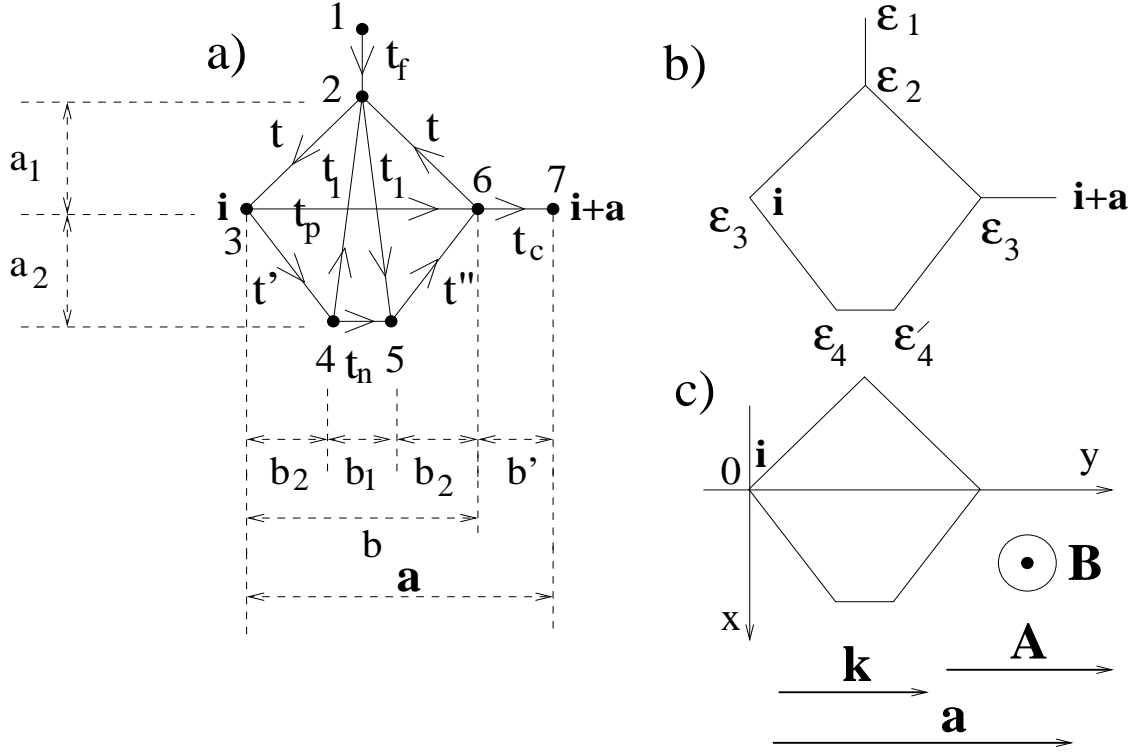}} 
\caption{a) The pentagon unit cell of the system under study. The quantities
$a_1,a_2,b=2b_2+b_1, a=|{\bf a}|=b+b'$ characterize the geometrical extension
of the pentagon cell. The hopping matrix elements along different bonds are
given by $t,t',t",t_f,t_c,t_n,t_p,t_1$, while the arrows on the bonds show
the hopping direction for which the Peierls phase factor $\phi_{i_1,i_2},
{\bf i}_1 \to {\bf i}_2$ has been calculated. One notes that ${\bf i}$ is the
lattice site where the presented pentagon cell is placed, ${\bf a}$ is the
unique Bravais vector of the system, and different sites of the unit cell
placed at the site ${\bf i}$ are denoted by ${\bf i}+{\bf r}_{\alpha}$. The
sublattice index is $\alpha=1,2,...,6$, while for mathematical convenience
${\bf r}_3=0$.
b) The on-site one-particle potentials $\epsilon_{\alpha}$ at different sites
of the pentagon, $\alpha=1,2,...,6$. c) The system of coordinates xOy 
used during the calculation of the
Peierls phase factors $\phi_{i,j}$. The direction of the 
external magnetic field is presented by
${\bf B}$, ${\bf A}$ shows the direction of
the corresponding vector potential ${\bf A}$, the unique Bravais vector 
direction is
presented by ${\bf a}$, while the direction of the wave vector 
is shown by ${\bf k}$. }      
\end{figure}                                                               

\begin{eqnarray}
\phi_{i,j}=\frac{2\pi}{\Phi_0}\int_i^j {\bf A}{\bf dl},
\label{UE1}
\end{eqnarray}
where $\Phi_0=hc/e$ represents the flux quantum. We must underline that if one
moves in positive trigonometric direction an electron starting from the site
${\bf i}_1$, around a closed path C containing the sites ${\bf i}_1, {\bf i}_2,
{\bf i}_3, {\bf i}_4, ..., {\bf i}_n, {\bf i}_1$, hence turning back to the 
starting point ${\bf i}_1$, one obtains
\begin{eqnarray}
\phi_{{\bf i}_1,{\bf i}_1}&=&\frac{2\pi}{\Phi_0}[\int_{{\bf i}_1}^{{\bf i}_2} {\bf A} 
d {\bf l} + \int_{{\bf i}_2}^{{\bf i}_3} {\bf A} d {\bf l} + \int_{{\bf i}_3}^{{\bf i}_4} 
{\bf A} d {\bf l} +...+\int_{{\bf i}_n}^{{\bf i}_1} {\bf A} d {\bf l} ] =
\frac{2\pi}{\Phi_0} \oint_C {\bf A} d {\bf l}  =\frac{2\pi}{\Phi_0} \int_S
(\nabla \times {\bf A}) d {\bf S} 
\nonumber\\
&=& \frac{2\pi}{\Phi_0} \int_S {\bf B} d{\bf S}
=\frac{2\pi}{\Phi_0} \Phi,
\label{UE2}
\end{eqnarray}
where $\Phi$ is the magnetic flux through the surface S enclosed by the path
C. In Eq.(\ref{UE2}), at the third step the Stoke's theorem, at the fourth step
the relation between the magnetic induction ${\bf B}$ and the vector potential
${\bf A}$, namely ${\bf B}=\nabla \times {\bf A}$, and at the fifth step the
definition of the magnetic flux $\Phi=\int_S {\bf B} d {\bf S}$ has been used.
As a consequence of Eq.(\ref{UE2}), if one adds the Peierls phase factors
$\phi_{{\bf i}_1,{\bf i}_2}$ corresponding to the bond $b_{{\bf i}_1,{\bf i}_2}=
({\bf i}_1,{\bf i}_2)$ along bonds enclosing an arbitrary surface S with a 
closed path C, one must reobtain the flux $\Phi$ threading the surface S, 
multiplied by $2\pi/\Phi_0$, namely (see also Fig.2)

\begin{figure} [h]                                                         
\centerline{\includegraphics[width=4cm,height=4cm]{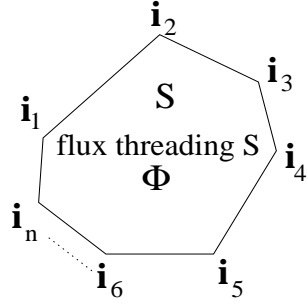}} 
\caption{The sum of Peierls phase factors along a closed path enclosing
the surface S gives back the flux threading the surface S, see Eq.(\ref{UE3}).
}      
\end{figure}                                                               

\begin{eqnarray}
\phi_{{\bf i}_1,{\bf i}_2}+\phi_{{\bf i}_2,{\bf i}_3}+\phi_{{\bf i}_3,{\bf i}_4}+ ... +
\phi_{{\bf i}_n,{\bf i}_1}= \frac{2\pi}{\Phi_0} \Phi
\label{UE3}
\end{eqnarray}

The Eq.(\ref{UE3}) can be used for checking the calculations providing the
Peierls phase factors $\phi_{{\bf i}_1,{\bf i}_2}$.

The calculation of the Peierls phase factors is performed in details in 
Appendix A. The obtained (and see (\ref{UE11}), checked) result is
\begin{eqnarray}
\delta &=& \frac{2\pi}{\Phi_0}B \frac{a_1b}{4}=\phi_{6,2}=\phi_{2,3}, \quad
\delta'= \frac{2\pi}{\Phi_0}B \frac{a_2b_2}{2}=\phi_{3,4}=\phi_{5,6},
\nonumber\\
\delta_1 &=& \frac{2\pi}{\Phi_0}B \frac{b_1(a_2-a_1)}{4}=\phi_{4,2}=\phi_{2,5},
\quad \delta_n =\frac{2\pi}{\Phi_0}B a_2 b_1=\phi_{4,5},
\nonumber\\
0 &=& \phi_{3,6}=\phi_{6,7}=\phi_{1,2}.
\label{UE10}
\end{eqnarray}

\subsection{The effect of the electric field}

The external electric potential is taken into consideration by producing
modifications in the on-site one particle potentials ($\epsilon_{\bf j}$)
in a site selective manner. This means that the same $\Delta \epsilon$
variation appears in the on-site one particle potential in each cell at 
the same site, i.e. for an arbitrary site ${\bf j}$, given by the field, 
$\epsilon_{{\bf j}+n {\bf a}}+\Delta \epsilon$
appears for each integer $n$, where ${\bf a}$ is the unique Bravais vector
of the system. 

We mention that for the site selective modification of the
local on-site one-particle potentials different techniques can be used
in practice. In order to exemplify, we mention:
one dimensional FETs (field effect transistors) disposed in an 
array geometry \cite{Fet1,Fet2}, directed electric fields
applied on short distances \cite{Fet3}, charge transfer through site-selective 
interaction \cite{Fet4}, site-selective conjugation into the polymer
\cite{Fet5}, site selective binding to the polymer \cite{Fet6},  
site-selective application of potential during bipolar electrolysis
\cite{Fet7}, or even site selective introduction of functional groups into the
polymer \cite{Fet8}.

At the level of the description, we simply use the notation $\epsilon_{\bf j}$
for the on-site potentials in the Hamiltonian, keeping in mind that this
value, if needed, can be modified in a site selective manner as mentioned
above.

\section{The Hamiltonian of the system in the presented 
conditions}

\subsection{The Hamiltonian}

Taking into account the Peierls phase factors from Eq.(\ref{UE10}), and the
presence of the site selective on-site one particle potentials, one has
$\hat H=\hat H_{kin,1}+\hat H_{kin,2}+\hat H_{int}$, where
\begin{eqnarray}
\hat H_{kin,1} &=& \sum_{\sigma} \sum_{i=1}^{N_c} \big[
(t e^{i\delta} \hat c^{\dagger}_{{\bf i}+{\bf r_2},\sigma} \hat c_{{\bf i}+
{\bf r}_6,\sigma} + t e^{i\delta} \hat c^{\dagger}_{{\bf i}+{\bf r_3},\sigma} 
\hat c_{{\bf i}+{\bf r}_2,\sigma}) + 
(t' e^{i\delta'} \hat c^{\dagger}_{{\bf i}+{\bf r_4},\sigma} \hat c_{{\bf i}+
{\bf r}_3,\sigma} + t" e^{i\delta'} \hat c^{\dagger}_{{\bf i}+{\bf r_6},\sigma} 
\hat c_{{\bf i}+{\bf r}_5,\sigma}) 
\nonumber\\
&+& 
(t_1 e^{i\delta_1} \hat c^{\dagger}_{{\bf i}+{\bf r_2},\sigma} \hat c_{{\bf i}+
{\bf r}_4,\sigma} + t_1 e^{i\delta_1} \hat c^{\dagger}_{{\bf i}+{\bf r_5},
\sigma} \hat c_{{\bf i}+{\bf r}_2,\sigma}) +
t_n e^{i\delta_n} \hat c^{\dagger}_{{\bf i}+{\bf r_5},\sigma} \hat c_{{\bf i}+
{\bf r}_4,\sigma} 
\nonumber\\
&+& 
(t_c \hat c^{\dagger}_{{\bf i}+{\bf r_7},\sigma} \hat c_{{\bf i}+{\bf r}_6,
\sigma} + t_p \hat c^{\dagger}_{{\bf i}+{\bf r_6},\sigma} \hat c_{{\bf i},
\sigma} + t_f \hat c^{\dagger}_{{\bf i}+{\bf r_2},\sigma} \hat c_{{\bf i}+
{\bf r}_1,\sigma}) + H.c. \big],
\nonumber\\
\hat H_{kin,2}&=&\sum_{\sigma} \sum_{i=1}^{N_c} \big[\epsilon_1 \hat n_{{\bf i}+{\bf
r}_1,\sigma} + \epsilon_2 \hat n_{{\bf i}+{\bf r}_2,\sigma} +
\epsilon_3 (\hat n_{{\bf i}+{\bf r}_3,\sigma} +\hat n_{{\bf i}+{\bf r}_6,\sigma}) +
\epsilon_4 (\hat n_{{\bf i}+{\bf r}_4,\sigma} +\hat n_{{\bf i}+{\bf r}_5,\sigma}) \big],
\nonumber\\
\hat H_{int} &=&\sum_{i=1}^{N_c} \big[U_1 \hat n_{{\bf i}+{\bf r}_1,\uparrow}
\hat n_{{\bf i}+{\bf r}_1,\downarrow}+U_2 \hat n_{{\bf i}+{\bf r}_2,\uparrow}
\hat n_{{\bf i}+{\bf r}_2,\downarrow}+U_3 (\hat n_{{\bf i}+{\bf r}_3,\uparrow}
\hat n_{{\bf i}+{\bf r}_3,\downarrow}+\hat n_{{\bf i}+{\bf r}_6,\uparrow}
\hat n_{{\bf i}+{\bf r}_6,\downarrow})
\nonumber\\
&+& U_4 (\hat n_{{\bf i}+{\bf r}_4,\uparrow}
\hat n_{{\bf i}+{\bf r}_4,\downarrow}+\hat n_{{\bf i}+{\bf r}_5,\uparrow}
\hat n_{{\bf i}+{\bf r}_5,\downarrow}) \big].
\label{UE12}
\end{eqnarray}
We append the following observations relating the used Hamiltonian presented
above. a) The studied system is in fact a conducting polymer free of magnetic
atoms, hence (\ref{UE12}) describes itinerant carriers holding the same charge.
Hence the inter-carrier interaction is of Coulomb type. b) The studied chain
is in fact a many-body itinerant system, hence a strong screening of the Coulomb
interaction is present. Because of this reason the longer tails of the Coulomb
repulsion are neglected, and the inter-carrier interaction in (\ref{UE12})
appears as on-site Coulomb repulsion, whose strength at the site ${\bf j}$
is given by $U_{\bf j}$ for an arbitrary ${\bf j}+n {\bf a}$, where $n$ is an 
arbitrary integer (i.e. $U_{\bf j}$ is fixed for a given type of site in each 
cell). c) The carriers created by the $\hat c^{\dagger}_{{\bf j},\sigma}$ canonical
Fermi operators can be considered in fact fermionic quasiparticles, so their 
mobility and effective mass must not coincide with the bare electron mobility 
and effective mass. In this view, (\ref{UE12}) can be considered as an
effective Hamiltonian in which even polaronic concepts are in a given extent
incorporated. Further electron-phonon interactions are neglected in the starting
Hamiltonian. These neglected interactions become important at system half 
filling, but away from this carrier concentration (i.e. electron or hole doped
polymer) their importance is small \cite{Intr8,Intr25,Intr25a}.

\subsection{The Hamiltonian in momentum space}

By transforming the $\hat H_{kin}= \hat H_{kin,1}+\hat H_{kin,2}$ expression in 
${\bf k}$-space we use: i) $\hat c_{{\bf i}+{\bf r}_n,\sigma}=(1/\sqrt{N_c})
\sum_{{\bf k}=1}^{N_c}e^{-i{\bf k}({\bf i}+{\bf r}_n)} \hat c_{n,{\bf k},\sigma}$, where
$n=1,2,...,6$ is the sublattice index, ii) $(1/N_c) \sum_{{\bf i}=1}^{N_c}
e^{i({\bf k}_1-{\bf k}_2){\bf i}}=\delta_{{\bf k}_1-{\bf k}_2}$,
one obtains
\begin{eqnarray}
&&\hat H_{kin}=\sum_{{\bf k},\sigma}\Big[
t_c(e^{+i{\bf k}({\bf a}-{\bf r}_6)}\hat c^{\dagger}_{3,{\bf k},\sigma}
\hat c_{6,{\bf k},\sigma}+e^{-i{\bf k}({\bf a}-{\bf r}_6)}
\hat c^{\dagger}_{6,{\bf k},\sigma}\hat c_{3,{\bf k},\sigma})+
t_p(e^{+i{\bf k}{\bf a}}\hat c^{\dagger}_{3,{\bf k},\sigma}
\hat c_{3,{\bf k},\sigma}+e^{-i{\bf k}{\bf a}}
\hat c^{\dagger}_{3,{\bf k},\sigma}\hat c_{3,{\bf k},\sigma})+
\nonumber\\
&&+t(e^{i\delta} e^{i{\bf k}({\bf r}_2-{\bf r}_6)}\hat c^{\dagger}_{2,{\bf k},\sigma}
\hat c_{6,{\bf k},\sigma}+e^{-i\delta} e^{-i{\bf k}({\bf r}_2-{\bf r}_6)}
\hat c^{\dagger}_{6,{\bf k},\sigma}\hat c_{2,{\bf k},\sigma})+
t(e^{i\delta} e^{-i{\bf k}{\bf r}_2}\hat c^{\dagger}_{3,{\bf k},\sigma}
\hat c_{2,{\bf k},\sigma}+e^{-i\delta} e^{+i{\bf k}{\bf r}_2}
\hat c^{\dagger}_{2,{\bf k},\sigma}\hat c_{3,{\bf k},\sigma})
\nonumber\\
&&+
t'(e^{i\delta'} e^{+i{\bf k}{\bf r}_4}\hat c^{\dagger}_{4,{\bf k},\sigma}
\hat c_{3,{\bf k},\sigma}+e^{-i\delta'} e^{-i{\bf k}{\bf r}_4}
\hat c^{\dagger}_{3,{\bf k},\sigma}\hat c_{4,{\bf k},\sigma})+
t"(e^{i\delta'} e^{+i{\bf k}({\bf r}_6-{\bf r}_5)}\hat c^{\dagger}_{6,{\bf k},\sigma}
\hat c_{5,{\bf k},\sigma}+e^{-i\delta'} e^{-i{\bf k}({\bf r}_6-{\bf r}_5)}
\hat c^{\dagger}_{5,{\bf k},\sigma}\hat c_{6,{\bf k},\sigma})
\nonumber\\
&&+
t_1(e^{i\delta_1} e^{+i{\bf k}({\bf r}_2-{\bf r}_4)}\hat c^{\dagger}_{2,{\bf k},\sigma}
\hat c_{4,{\bf k},\sigma}+e^{-i\delta_1} e^{-i{\bf k}({\bf r}_2-{\bf r}_4)}
\hat c^{\dagger}_{4,{\bf k},\sigma}\hat c_{2,{\bf k},\sigma})+
t_1(e^{i\delta_1} e^{+i{\bf k}({\bf r}_5-{\bf r}_2)}\hat c^{\dagger}_{5,{\bf k},\sigma}
\hat c_{2,{\bf k},\sigma}
\nonumber\\
&&+e^{-i\delta_1} e^{-i{\bf k}({\bf r}_5-{\bf r}_2)}
\hat c^{\dagger}_{2,{\bf k},\sigma}\hat c_{5,{\bf k},\sigma})+
t_n(e^{i\delta_n} e^{+i{\bf k}({\bf r}_5-{\bf r}_4)}\hat c^{\dagger}_{5,{\bf k},\sigma}
\hat c_{4,{\bf k},\sigma}+e^{-i\delta_n} e^{-i{\bf k}({\bf r}_5-{\bf r}_4)}
\hat c^{\dagger}_{4,{\bf k},\sigma}\hat c_{5,{\bf k},\sigma})
\nonumber\\
&&+
t_f(e^{+i{\bf k}({\bf r}_2-{\bf r}_1)}\hat c^{\dagger}_{2,{\bf k},\sigma}
\hat c_{1,{\bf k},\sigma}+e^{-i{\bf k}({\bf r}_2-{\bf r}_1)}
\hat c^{\dagger}_{1,{\bf k},\sigma}\hat c_{2,{\bf k},\sigma})+
\epsilon_1\hat c^{\dagger}_{1,{\bf k},\sigma}\hat c_{1,{\bf k},\sigma}+
\epsilon_2\hat c^{\dagger}_{2,{\bf k},\sigma}\hat c_{2,{\bf k},\sigma}
\nonumber\\
&&+
\epsilon_3(\hat c^{\dagger}_{3,{\bf k},\sigma}\hat c_{3,{\bf k},\sigma}+
\hat c^{\dagger}_{6,{\bf k},\sigma}\hat c_{6,{\bf k},\sigma})+
\epsilon_4\hat c^{\dagger}_{4,{\bf k},\sigma}\hat c_{4,{\bf k},\sigma}+
\epsilon'_4\hat c^{\dagger}_{5,{\bf k},\sigma}\hat c_{5,{\bf k},\sigma} \Big].
\label{UE13}
\end{eqnarray}
Now taking into consideration that ${\bf k}$ is directed along the line of
the chain (the line connecting the sites ${\bf i}$ and ${\bf i}+{\bf a}$,
hence ${\bf k}$ and ${\bf a}$ are parallel), and $k={\bf a}{\bf k}=2m\pi/N_c$
where $m=0,1,2,...,N_c-1$, one has (see Fig.1a):
\begin{eqnarray}
&&{\bf k}({\bf a}-{\bf r}_6)=kb', \quad {\bf k}{\bf a}=k, \quad
{\bf k}({\bf r}_2-{\bf r}_6)=-\frac{k b}{2}, \quad {\bf k}{\bf r}_2=
\frac{k b}{2}, \quad
{\bf k}({\bf r}_5-{\bf r}_4)=kb_1, \quad {\bf k}({\bf r}_2-{\bf r}_1)=0,
\nonumber\\
&&{\bf k}{\bf r}_4=kb_2, \quad {\bf k}({\bf r}_6-{\bf r}_5)=kb_2, \quad
{\bf k}({\bf r}_2-{\bf r}_4)=k(\frac{b}{2}-b_2), 
\nonumber\\
&&{\bf k}({\bf r}_5-{\bf r}_2)=k(b_1+b_2-\frac{b}{2})=k\frac{b_1}{2}
=k(\frac{b}{2}-b_2),
\label{UE14}
\end{eqnarray}
consequently the $\hat H_{kin}$ from (\ref{UE13}) becomes 
\begin{eqnarray}
&&\hat H_{kin}=\sum_{{\bf k},\sigma} \Big[
t_c(e^{+ikb'}\hat c^{\dagger}_{3,{\bf k},\sigma}\hat c_{6,{\bf k},\sigma}+
e^{-ikb'}\hat c^{\dagger}_{6,{\bf k},\sigma}\hat c_{3,{\bf k},\sigma})+
(2t_p \cos k +\epsilon_3)\hat c^{\dagger}_{3,{\bf k},\sigma}\hat c_{3,{\bf k},\sigma}
\nonumber\\
&&+t(e^{i\delta} e^{-i \frac{kb}{2}}\hat c^{\dagger}_{2,{\bf k},\sigma}
\hat c_{6,{\bf k},\sigma}+e^{-i\delta} e^{+i \frac{kb}{2}}
\hat c^{\dagger}_{6,{\bf k},\sigma}\hat c_{2,{\bf k},\sigma})+
t(e^{i\delta} e^{-i \frac{kb}{2}} \hat c^{\dagger}_{3,{\bf k},\sigma}
\hat c_{2,{\bf k},\sigma}+e^{-i\delta} e^{+i \frac{kb}{2}}
\hat c^{\dagger}_{2,{\bf k},\sigma}\hat c_{3,{\bf k},\sigma})
\nonumber\\
&&+
t'(e^{i\delta'} e^{+ikb_2}\hat c^{\dagger}_{4,{\bf k},\sigma}
\hat c_{3,{\bf k},\sigma}+e^{-i\delta'} e^{-ikb_2}
\hat c^{\dagger}_{3,{\bf k},\sigma}\hat c_{4,{\bf k},\sigma})+
t"(e^{i\delta'} e^{+ikb_2}\hat c^{\dagger}_{6,{\bf k},\sigma}
\hat c_{5,{\bf k},\sigma}+e^{-i\delta'} e^{-ikb_2}
\hat c^{\dagger}_{5,{\bf k},\sigma}\hat c_{6,{\bf k},\sigma})
\nonumber\\
&&+
t_1(e^{i\delta_1} e^{+ik(\frac{b}{2}-b_2)}\hat c^{\dagger}_{2,{\bf k},\sigma}
\hat c_{4,{\bf k},\sigma}+e^{-i\delta_1} e^{-ik(\frac{b}{2}-b_2)}
\hat c^{\dagger}_{4,{\bf k},\sigma}\hat c_{2,{\bf k},\sigma})+
t_1(e^{i\delta_1} e^{+ik (\frac{b}{2}-b_2)}\hat c^{\dagger}_{5,{\bf k},\sigma}
\hat c_{2,{\bf k},\sigma}
\nonumber\\
&&+
e^{-i\delta_1}e^{-ik(\frac{b}{2}-b_2)}\hat c^{\dagger}_{2,{\bf k},
\sigma} \hat c_{5,{\bf k},\sigma})+
t_n(e^{i\delta_n} e^{+ikb_1}\hat c^{\dagger}_{5,{\bf k},\sigma}
\hat c_{4,{\bf k},\sigma}+e^{-i\delta_n} e^{-ikb_1}
\hat c^{\dagger}_{4,{\bf k},\sigma}\hat c_{5,{\bf k},\sigma})
\nonumber\\
&&+
t_f(\hat c^{\dagger}_{2,{\bf k},\sigma}\hat c_{1,{\bf k},\sigma}+
\hat c^{\dagger}_{1,{\bf k},\sigma}\hat c_{2,{\bf k},\sigma})
+\epsilon_1\hat c^{\dagger}_{1,{\bf k},\sigma}\hat c_{1,{\bf k},\sigma}+
\epsilon_2\hat c^{\dagger}_{2,{\bf k},\sigma}\hat c_{2,{\bf k},\sigma}+
\epsilon_3 \hat c^{\dagger}_{6,{\bf k},\sigma}\hat c_{6,{\bf k},\sigma}
\nonumber\\
&&+
\epsilon_4\hat c^{\dagger}_{4,{\bf k},\sigma}\hat c_{4,{\bf k},\sigma}+
\epsilon'_4\hat c^{\dagger}_{5,{\bf k},\sigma}\hat c_{5,{\bf k},\sigma} \Big].
\label{UE15}
\end{eqnarray}
Introducing the $(6 \times 1)$ column vector $\hat C$ composed in 
order 
from the components $\hat c_{1,{\bf k},\sigma}, \hat c_{2,{\bf k},\sigma},
\hat c_{3,{\bf k},\sigma},...,\hat c_{6,{\bf k},\sigma}$, whose transposed adjoint
provides the $(1\times 6)$ row vector 
${\hat C}^{\dagger}=(\hat c^{\dagger}_{1,{\bf k},\sigma},
\hat c^{\dagger}_{2,{\bf k},\sigma}, \hat c^{\dagger}_{3,{\bf k},\sigma},...,
\hat c^{\dagger}_{6,{\bf k},\sigma})$, the $\hat H_{kin}$ operator can be written 
as
\begin{eqnarray}
\hat H_{kin}=\sum_{{\bf k},\sigma} {\hat C}^{\dagger} {\tilde M} \hat C,
\label{UE16}
\end{eqnarray}
where the $(6 \times 6)$ matrix $\tilde M$ has the form
\begin{eqnarray}
\tilde M =
\left( \begin{array}{cccccc}
\epsilon_1 & t_f & 0 & 0 & 0 & 0 \\
t_f & \epsilon_2 & t e^{-i\delta}e^{+ i \frac{kb}{2}} & t_1e^{i\delta_1}e^{ik(
\frac{b}{2}-b_2)} & t_1e^{-i\delta_1}e^{-ik(\frac{b}{2}-b_2)} 
& t e^{i\delta}e^{-i \frac{k b}{2}} \\
0 & t e^{i\delta}e^{-i \frac{k b}{2} } & 2t_p \cos k+\epsilon_3 & t'e^{-i\delta'} 
e^{-i k b_2} & 0 & t_c e^{i k b'} \\
0 & t_1e^{-i\delta_1}e^{-ik(\frac{b}{2}-b_2)}
& t'e^{i\delta'} e^{+ i k b_2} & \epsilon_4 & t_ne^{-i\delta_n} e^{- i k b_1} & 0 \\
0 & t_1e^{i\delta_1}e^{ik(\frac{b}{2}-b_2)} & 0 & t_n e^{i\delta_n}e^{+ i k b_1} & 
\epsilon'_4 & t"e^{-i\delta'} e^{- i k b_2} \\
0 &  te^{-i\delta} e^{+i \frac{k b}{2}} & t_c e^{- i k b'} & 0 & t"e^{i\delta'} 
e^{+ i k b_2} & \epsilon_3
\end{array} \right) .
\label{UE17}
\end{eqnarray}

\subsection{The bare band structure}

The band structure is obtained from the secular equation of $\tilde M$,
hence $Det[\tilde M -\lambda \tilde 1]=0$, where $\tilde 1$ is the 
$(6\times 6)$ unity matrix, and $\lambda$ provides the eigenvalues.
Introducing the notations
\begin{eqnarray}
&&E_1=\epsilon_1-\lambda, \quad E_2=\epsilon_2-\lambda, \quad
E_3=\epsilon_3-\lambda, \quad E'_3=\epsilon_3+2t_p \cos k -\lambda,
\nonumber\\
&&E_4=\epsilon_4-\lambda, \quad E'_4=\epsilon_5-\lambda=\epsilon'_4-\lambda,
\label{UE18}
\end{eqnarray}
the secular equation for $\tilde M$ gives as eigenvalues the
$\lambda$ solutions of the following equation:
\begin{eqnarray}
&&(E_1E_2-t_f^2)\Big[E'_3(E_3E_4E'_4-E_4{t"}^2-E_3t_n^2)-{t'}^2(E_3E'_4-{t"}^2)
-t_c^2(E_4E'_4-t_n^2)
\nonumber\\
&&-2t't"t_nt_c \cos [k+(\delta_n+2\delta')] \Big]+
\Big[E_1E'_4(t_1^2t_c^2+t^2{t'}^2)+E_1E_4(t_1^2t_c^2+t^2{t"}^2)
-E_1E'_3t_1^2(E_3E'_4-{t"}^2)
\nonumber\\
&&-E_1E'_3t^2(E_4E'_4-t_n^2)
-E_1E_3t_1^2(E_4E'_3-{t'}^2)-E_1E_3t^2(E_4E'_4-t_n^2) \Big]
\nonumber\\
&&+2E_1t_ct^2(E_4E'_4-t_n^2)\cos(k-2\delta)+2E_1t^2t't"t_n
\cos(2\delta+2\delta'+\delta_n)+2E_1t_1^2t_ct't"\cos[k+(2\delta'+2\delta_1)]
\nonumber\\
&&-2E_1t_1t_ct(E'_4t'+E_4t")\cos[k+(\delta_1+\delta'-\delta)]+
+2E_1t_1t_ct_nt(t'+t")\cos[k+(\delta'+\delta_n-\delta-\delta_1)]
\nonumber\\
&&-2E_1t_1t_nt(E_3t'+E'_3t")\cos(\delta+\delta'+\delta_n-\delta_1)+
2E_1tt_1[t'(E_3E'_4-{t"}^2)+t"(E_4E'_3-{t'}^2)]\cos(\delta+\delta'+\delta_1)
\nonumber\\
&&+2E_1t_1^2t_n(E_3E'_3-t_c^2)\cos(\delta_n-2\delta_1) = 0.
\label{UE19}
\end{eqnarray}  
From this equation results that at $t_1=0$ the band structure is obtained
from the equation
\begin{eqnarray}
&&(E_1E_2-t_f^2)\Big[E'_3(E_3E_4E'_4-E_4{t"}^2-E_3t_n^2)-{t'}^2(E_3E'_4-{t"}^2)
-t_c^2(E_4E'_4-t_n^2)
\nonumber\\
&&-2t't"t_nt_c \cos [k+(\delta_n+2\delta')] \Big]+
t^2 E_1 \Big[E'_4{t'}^2+E_4{t"}^2
-E'_3(E_4E'_4 -t_n^2)-E_3(E_4E'_4-t_n^2) \Big]
\nonumber\\
&&+2E_1t^2 \Big[ t_c(E_4E'_4-t_n^2)\cos(k-2\delta)+t't"t_n
\cos(2\delta+2\delta'+\delta_n)\Big]= 0.
\label{UE20}
\end{eqnarray}  
Now if $t_p=0$ and because of this reason $E_3=E'_3$, furthermore $t"=t'$ and
$E'_4=E_4$, equation (\ref{UE20}) becomes
\begin{eqnarray}
&&(E_1E_2-t_f^2)\Big[E_3E_4(E_3E_4-2{t'}^2)+{t'}^4
-t_c^2(E_4^2-t_n^2)-t_n^2E_3^2
-2{t'}^2t_nt_c \cos [k+(\delta_n+2\delta')] \Big]
\nonumber\\
&&+2t^2 E_1 \Big[E_4{t'}^2-E_3(E_4^2-t_n^2)+
t_c(E^2_4-t_n^2)\cos(k-2\delta)+{t'}^2t_n
\cos(2\delta+2\delta'+\delta_n)\Big]= 0.
\label{UE21}
\end{eqnarray}  
 Without external magnetic field, Eq.(\ref{UE21}) reads
\begin{eqnarray}
&&2t_c[-t_n{t'}^2(E_1E_2-t_f^2)+t^2E_1(E_4^2-t_n^2)]\cos k +
(E_1E_2-t_f^2)[(E_3E_4-{t'}^2)^2-t_c^2(E_4-t_n^2)-t_n^2E_3^2]+
\nonumber\\
&&2t^2E_1[{t'}^2(E_4+t_n)-E_3(E_4^2-t_n^2)]=0.
\label{UE22}
\end{eqnarray}
This last equation, at $t'=t$ and $E_1=E_2$ gives back the old
result known for the pentagon chains \cite{Intr8}
\begin{eqnarray}
&&2t_ct^2\{ E_1(E_4^2-t_n^2)-t_n(E_1^2-t_f^2)\} \cos k +
(E_1^2-t_f^2) [(E_3E_4-t^2)^2-t_n^2E_3^2-t_c^2E_4^2+t_n^2t_c^2]+
\nonumber\\
&&2t^2E_1[E_3(t_n^2-E_4^2)+t^2(E_4+t_n)]=0.
\label{UE23}
\end{eqnarray}

The study of (\ref{UE19}), or the comparison between
e.g. (\ref{UE22}) and (\ref{UE21}) shows that the external fields
have a substantial effect even at the level of the bare band structure
modifying the physical properties of the system.
These modifications, besides the effective mass and carrier average velocity
changes that they automatically produce, can lead to band flattening effects 
or modifications of proximities to flat bands (see for importance e.g. Refs.
\cite{Intr8,Intr25,Intr25a,Rich1,Rich2,Rich3}), or level crossing effects
(qualitative changes produced by this effects are exemplified e.g. by
Refs. \cite{LCR1,LCR2,LCR3,LCR4,LCR5,LCR6,LCR7}).

\section{Exact ground states in the interacting case. I.}

The modifications in the bare band structure presented in the previous
section lead also to modifications of the phase diagram of the system
described by the whole Hamiltonian.
This means e.g. that in several region of the parameter space new phases
appear. We exemplify this fact in the following two Sections for the case of
magnetic phases. The deduction of the many-body ground states will be made
in a non-approximated manner following the technique of the Hamiltonian
transformation in positive semidefinite form presented in the introductory
part (Sect.I.).
For simplicity, only the case of the symmetric cell with
nearest neighbor hoppings ($t_p=t_1=0$, 
$\epsilon_5=\epsilon_4$, $\epsilon_3=\epsilon_6$, $t'=t"$) will be analyzed
in details in Sections IV-V.

\subsection{The decomposition step}

We use 5 block operators for decomposition (see Fig.4)
for each cell. One has
\begin{eqnarray}
&& \hat A_{{\bf i},\sigma}= a_2 \hat c_{{\bf i}+{\bf r}_2,\sigma}+
 a_3 \hat c_{{\bf i}+{\bf r}_3,\sigma}+ a_4 \hat c_{{\bf i}+{\bf r}_4,\sigma},
\quad \hat F_{{\bf i},\sigma}=f_1 \hat c_{{\bf i}+{\bf r}_1,\sigma}+
f_2 \hat c_{{\bf i}+{\bf r}_2,\sigma},
\nonumber\\
&& \hat B_{{\bf i},\sigma}= b_2 \hat c_{{\bf i}+{\bf r}_2,\sigma}+
 b_4 \hat c_{{\bf i}+{\bf r}_4,\sigma}+ b_5 \hat c_{{\bf i}+{\bf r}_5,\sigma},
\quad \hat G_{{\bf i},\sigma}=g_6 \hat c_{{\bf i}+{\bf r}_6,\sigma}+
g_7 \hat c_{{\bf i}+{\bf a},\sigma},
\nonumber\\
&& \hat D_{{\bf i},\sigma}= d_2 \hat c_{{\bf i}+{\bf r}_2,\sigma}+
 d_5 \hat c_{{\bf i}+{\bf r}_5,\sigma}+ d_6 \hat c_{{\bf i}+{\bf r}_6,\sigma}.
\label{E2.1}
\end{eqnarray}

\begin{figure} [h]                                                         
\centerline{\includegraphics[width=6cm,height=5cm]{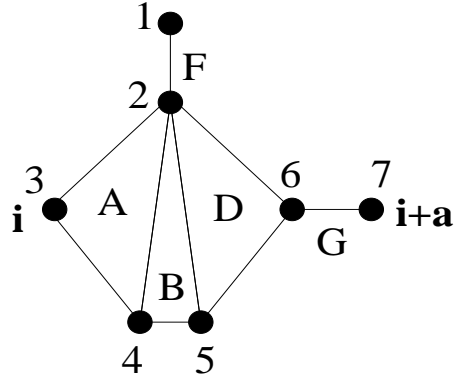}} 
\caption{The used five block operators. Three (A,B,D) are defined on
triangles, while two (F,G) on bonds.}
\end{figure}

The matching equations providing $\hat H = \hat O + C$ in the form
\begin{eqnarray}
&&\hat H - \hat H_{int}= \sum_{{\bf i},\sigma} [ \hat A^{\dagger}_{{\bf i},\sigma}
\hat A_{{\bf i},\sigma} + \hat B^{\dagger}_{{\bf i},\sigma}
\hat B_{{\bf i},\sigma} + \hat D^{\dagger}_{{\bf i},\sigma}
\hat D_{{\bf i},\sigma} + \hat F^{\dagger}_{{\bf i},\sigma}
\hat F_{{\bf i},\sigma} + \hat G^{\dagger}_{{\bf i},\sigma}
\hat G_{{\bf i},\sigma}] -p\hat N,
\nonumber\\
&&\hat O =\sum_{{\bf i},\sigma} [ \hat A^{\dagger}_{{\bf i},\sigma}
\hat A_{{\bf i},\sigma} + \hat B^{\dagger}_{{\bf i},\sigma}
\hat B_{{\bf i},\sigma} + \hat D^{\dagger}_{{\bf i},\sigma}
\hat D_{{\bf i},\sigma} + \hat F^{\dagger}_{{\bf i},\sigma}
\hat F_{{\bf i},\sigma} + \hat G^{\dagger}_{{\bf i},\sigma}
\hat G_{{\bf i},\sigma}] + \hat H_{int}, \: C= -p\hat N,
\label{E2.2}
\end{eqnarray}
becomes
\begin{eqnarray}
&&te^{i\delta}=a^*_3a_2=d^*_2d_6, \quad t'e^{i\delta'}=a^*_4a_3=d^*_6d_5,
\nonumber\\
&&t_1e^{i\delta_1}=a^*_2a_4+b^*_2b_4 = d^*_5d_2+b^*_5b_2, \quad
t_ne^{i\delta_n}=b^*_5b_4,
\nonumber\\
&&t_f=f^*_2f_1, \quad t_c=g^*_7g_6, \quad \epsilon_1+p=|f_1|^2,
\nonumber\\
&&\epsilon_2+p=|f_2|^2+|a_2|^2+|b_2|^2+|d_2|^2,
\nonumber\\
&&\epsilon_3+p=|a_3|^2+|g_7|^2=|d_6|^2+|g_6|^2,
\nonumber\\
&&\epsilon_4+p=|a_4|^2+|b_4|^2=|d_5|^2+|b_5|^2.
\label{E2.3}
\end{eqnarray}
Solving the matching equations, 8 parameters can be directly expressed
\begin{eqnarray}
&&a_2=\frac{te^{i\delta}}{a^*_3}, \quad a^*_4=\frac{t'e^{i\delta'}}{a_3}, \quad
d^*_2=\frac{te^{i\delta}}{d_6}, \quad d_5=\frac{t'e^{i\delta'}}{d^*_6}, 
\nonumber\\
&&b^*_5=\frac{t_ne^{i\delta_n}}{b_4}, \quad g^*_7=\frac{t_c}{g_6}, \quad
f_1=\sqrt{\epsilon_1+p}, \quad f_2=\frac{t_f}{\sqrt{\epsilon_1+p}},
\label{E2.4}
\end{eqnarray}
and the remaining 7 matching equations with unknowns $b_2,b_4,a_3,d_6,g_6,p$
become

\begin{eqnarray}
&&b^*_2b_4 = t_1e^{i\delta_1}-\frac{tt'e^{-i(\delta+\delta')}}{|a_3|^2}, \quad
\frac{b_2}{b_4}=\frac{1}{t_ne^{i\delta_n}}\Big(t_1e^{i\delta_1}-\frac{tt'
e^{-i(\delta+\delta')}}{|d_6|^2}\Big),
\nonumber\\
&&\epsilon_2+p=\frac{t^2_f}{\epsilon_1+p}+\frac{t^2}{|a_3|^2}+\frac{t^2}{|d_6|^2}
+ |b_2|^2, \quad \epsilon_1+p \geq 0,
\nonumber\\
&&\epsilon_3+p=|d_6|^2+|g_6|^2=|a_3|^2+\frac{t_c^2}{|g_6|^2},
\nonumber\\
&&\epsilon_4+p=|b_4|^2+\frac{{t'}^2}{|a_3|^2}=\frac{{t'}^2}{|d_6|^2}+
\frac{t_n^2}{|b_4|^2}.
\label{E2.5}
\end{eqnarray}
Now the first relation from (\ref{E2.5}) gives
\begin{eqnarray}
&&|b_2|^2=\frac{1}{t_ne^{i\delta_n}}\Big(t_1e^{i\delta_1}-\frac{
tt'e^{-i(\delta+\delta')}}{|a_3|^2}\Big)\Big(t_1e^{i\delta_1}-\frac{tt'
e^{-i(\delta+\delta')}}{|d_6|^2}\Big),
\nonumber\\
&&|b_4|^2=t_ne^{-i\delta_n} \frac{t_1e^{i\delta_1}-\frac{tt'e^{-i(\delta+\delta')}}{
|a_3|^2}}{t_1e^{-i\delta_1}-\frac{tt'e^{+i(\delta+\delta')}}{|d_6|^2}}.
\label{E2.6}
\end{eqnarray}
For the case $t_1=0$ the equalities (\ref{E2.6}) provide
\begin{eqnarray}
&&|b_2|^2=\frac{t^2{t'}^2}{|a_3|^2|d_6|^2}\frac{1}{t_ne^{i[\delta_n+2(\delta+\delta')]}
},
\nonumber\\
&&|b_4|^2=\frac{|d_6|^2}{|a_3|^2}t_ne^{-i[\delta_n+2(\delta+\delta')]},
\label{E2.7}
\end{eqnarray}
from where is seen that $\bar t_n=t_ne^{i[\delta_n+2(\delta+\delta')]} \geq 0$ must 
hold. This means that at $t_1=0$, with $m$ as an arbitrary integer, one has
\begin{eqnarray}
&&t_n > 0, \quad \delta_n+2(\delta+\delta') = 2m\pi,
\nonumber\\
&&t_n < 0, \quad \delta_n+2(\delta+\delta') = (2m+1)\pi.
\label{E2.8}
\end{eqnarray}

\subsection{The solution of the matching equations}

Based on (\ref{E2.5},\ref{E2.7}) a simple solution of the matching equations
is obtained. The starting point in deducing it is the $|a_3|=|d_6|$
relation which is provided by Eqs. (\ref{E2.3}), see Appendix B.
Based on this equality one directly finds
$b_4=\sqrt{\bar t_n}, b_5=t_ne^{-i\delta_n}
/\sqrt{\bar t_n}$. In this case the last line of (\ref{E2.5}) is satisfied and
provides $|a_3|^2={t'}^2/(\epsilon_4+p-\bar t_n) > 0$. The $(\epsilon_3+p)$ 
equation gives a condition, namely $t_c^2=|g_6|^4$, and the $|g_6|$ value as
$|g_6|^2=\epsilon_3+p -{t'}^2/(\epsilon_4+p-\bar t_n) >0$. For $|b_2|$ one finds
from (\ref{E2.7}) that $|b_2|=(|tt'|/\sqrt{\bar t_n})(1/|a_3|^2)$. The equation
of $(\epsilon_2+p)$ determines the $p$ value. Summarizing the results one has
(note that $t_1=0$):
\begin{eqnarray}
&&a_3=\frac{|t'|}{\sqrt{\epsilon_4+p-\bar t_n}}, \quad
a_2=\frac{\sqrt{\epsilon_4+p-\bar t_n}}{|t'|}te^{i\delta}, \quad
a_4=\frac{\sqrt{\epsilon_4+p-\bar t_n}}{|t'|}t'e^{-i\delta'},
\nonumber\\
&&d_6=\frac{|t'|}{\sqrt{\epsilon_4+p-\bar t_n}}, \quad
d_2=\frac{\sqrt{\epsilon_4+p-\bar t_n}}{|t'|}te^{-i\delta}, \quad
d_5=\frac{\sqrt{\epsilon_4+p-\bar t_n}}{|t'|}t'e^{i\delta'},
\nonumber\\
&&b_4=\sqrt{\bar t_n}, \quad b_5=\frac{t_ne^{-i\delta_n}}{\sqrt{\bar t_n}},
\quad b_2=- \frac{t(\epsilon_4+p-\bar t_n)}{t'\sqrt{\bar t_n}}e^{i(
\delta+\delta')},
\nonumber\\
&&g_6=\sqrt{\frac{(\epsilon_3+p)(\epsilon_4+p-\bar t_n)-{t'}^2}{(\epsilon_4+p-
\bar t_n)}}, \quad g_7= t_c \sqrt{
\frac{(\epsilon_4+p-\bar t_n)}{(\epsilon_3+p)
(\epsilon_4+p-\bar t_n)-{t'}^2}},
\nonumber\\
&&f_1=\sqrt{\epsilon_1+p}, \quad f_2=\frac{t_f}{\sqrt{\epsilon_1+p}}.
\label{E2.9}
\end{eqnarray}
The solutions (\ref{E2.9}) are valid with the conditions
\begin{eqnarray}
&&|t_c|=\epsilon_3+p -\frac{{t'}^2}{(\epsilon_4+p-\bar t_n)} >0,
\nonumber\\
&&\epsilon_2+p=\frac{t_f^2}{\epsilon_1+p}+2\frac{t^2}{{t'}^2}(\epsilon_4+p-
\bar t_n) +\frac{t^2}{\bar t_n {t'}^2}(\epsilon_4+p-\bar t_n)^2 > 0,
\nonumber\\
&&\bar t_n > 0.
\label{E2.10}
\end{eqnarray}

\subsection{The deduction of the ground state}

By solving the matching equations in the previous subsection, the transformation
in (\ref{E2.2}) now becomes explicitly known, so we can start the deduction of
the ground state. The deduction procedure is based on constructing the
most general $|\Psi_g\rangle$ which satisfies $\hat O |\Psi_g\rangle=0$. The
technical steps followed in the presented case are described in extreme details
e.g. in Refs. \cite{Intr8,GX1}. Note that $|\Psi_g\rangle = \prod_{\alpha}
\hat B^{\dagger}_{\alpha} |0\rangle$ is considered, where $|0\rangle$ is the bare
Fock vacuum state with no carriers present, and $\alpha=({\bf i},\sigma)$ is a
condensed index.

\subsubsection{The possibility of a disconnected solution}

In this subsection we are looking for the possibility of $\hat B^{\dagger}_{\alpha}$
solutions in $|\Psi_g\rangle$ that do not touch each other, i.e. are defined
on disconnected geometrical blocks. This means that  $\hat B^{\dagger}_{\alpha}$
and $\hat B^{\dagger}_{\alpha'}$ for $\alpha \ne \alpha'$ do not have common 
operators, hence $ \{\hat B^{\dagger}_{\alpha}, \hat B^{\dagger}_{\alpha'} \}=0$
automatically holds.

For a disconnected solution, the operators from the wave vector have to have
the form (see Fig.4a)
\begin{eqnarray}
\hat B^{\dagger}_{{\bf i},\sigma} = x_1 \hat c^{\dagger}_{{\bf i}+{\bf r}_1,
\sigma}+x_2 \hat c^{\dagger}_{{\bf i}+{\bf r}_2,\sigma}+x_4 \hat c^{\dagger}_{
{\bf i}+{\bf r}_4,\sigma}+x_5 \hat c^{\dagger}_{{\bf i}+{\bf r}_5,\sigma}.
\label{E2.11}
\end{eqnarray}
The condition which determines the unknown coefficients $x_1,x_2,x_4,x_5$
is
\begin{eqnarray}
\{\hat A_{n,{\bf i},\sigma}, \hat B^{\dagger}_{{\bf j},\sigma'} \} =0,
\label{E2.11a}
\end{eqnarray}
which has to be satisfied  for all values of all indices. 
Here $\hat A_n$, for $n=1,2,...,5$ 
represents all block operators from (\ref{E2.1})
i.e. $\hat O$ from (\ref{E2.2}) becomes of the form
$\hat O=\sum_{n=1}^5 \hat A^{\dagger}_{n,{\bf i},\sigma}\hat A_{n,{\bf i},\sigma}+
\hat H_{int}$.
We underline that if (\ref{E2.11a}) is satisfied, it automatically
implies $\sum_{n=1}^5 \hat A^{\dagger}_{n,{\bf i},\sigma}\hat A_{n,{\bf i},\sigma}
|\Psi_g\rangle=0$. Starting from (\ref{E2.11a}) one finds
\begin{eqnarray}
&&x_1f_1+x_2f_2=0, \quad x_2a_2+x_4a_4=0,
\nonumber\\
&&x_2d_2+x_5d_5=0, \quad x_2b_2+x_4b_4+x_5b_5=0.
\label{E2.12}
\end{eqnarray}
Based on (\ref{E2.9}), from here one obtains
\begin{eqnarray}
 &&x_1=-\frac{t_f}{\epsilon_1+p}x_2, \quad x_4=-\frac{t}{t'}e^{i(\delta+\delta')}x_2,
\quad x_5=-\frac{t}{t'}e^{-i(\delta+\delta')}x_2,
\nonumber\\
&&\Big[-\frac{t(\epsilon_4+p-\bar t_n)}{t'\sqrt{\bar t_n}}e^{i(\delta+\delta')}
-\frac{t}{t'}\sqrt{\bar t_n}e^{i(\delta+\delta')}-\frac{t}{t'}\frac{t_n}{
\sqrt{\bar t_n}}e^{-i(\delta+\delta')}e^{-i\delta_n} \Big]x_2=0,
\label{E2.13}
\end{eqnarray}
which, taking into account the definition of $\bar t_n$ leads to
$\epsilon_4+p+\bar t_n=0$. This equality cannot be satisfied since 
$\epsilon_4+p >0$, and $\bar t_n >0$ hold. So disconnected ground state wave 
vector for this solution does not exist.

\begin{figure} [h]                                                         
\centerline{\includegraphics[width=14cm,height=5cm]{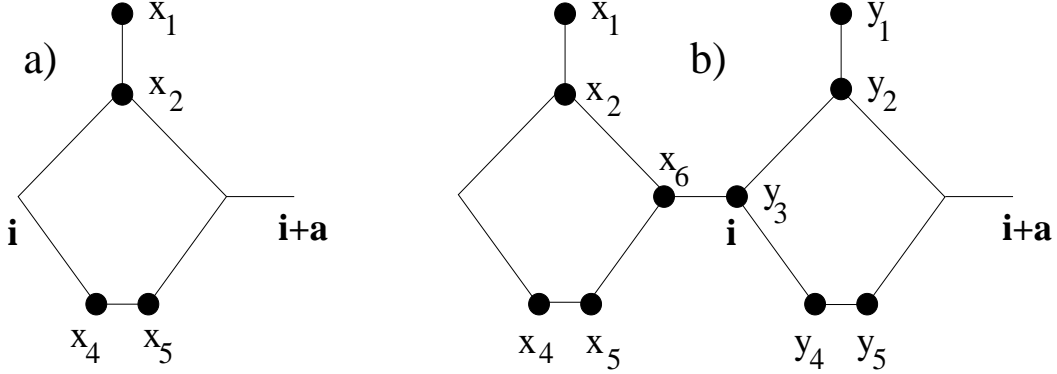}} 
\caption{The block operators in the ground state wave vector. a) disconnected
case, b) connected case. The dots show the sites that are belong to 
$\hat B^{\dagger}_{{\bf i},\sigma}$ together with the coefficients of the sites.}
\end{figure}

\subsubsection{Connected solution}

Contrary to the disconnected solution, now we are looking for 
$\hat B^{\dagger}_{\alpha}$ operators for $|\Psi_g\rangle$ with the
property that the $\{ \hat B^{\dagger}_{\alpha} \}$ manifold has 
$\alpha \ne \alpha'$ components which touch each other at least on one site,
i.e. two different $\hat B^{\dagger}_{\alpha}$ and $\hat B^{\dagger}_{\alpha'}$ for
some different $\alpha$ and $\alpha'$
have common creation operators.

The operators of the wave vector in the case of the connected solution
(see Fig.4b) have the expression
\begin{eqnarray}
\hat B^{\dagger}_{{\bf i},\sigma} &=& x_1 \hat c^{\dagger}_{{\bf i}-{\bf a}+
{\bf r}_1,\sigma} + x_2 \hat c^{\dagger}_{{\bf i}-{\bf a}+{\bf r}_2,\sigma} +
x_4 \hat c^{\dagger}_{{\bf i}-{\bf a}+{\bf r}_4,\sigma} + x_5 \hat c^{
\dagger}_{{\bf i}-{\bf a}+{\bf r}_5,\sigma} +x_6 \hat c^{\dagger}_{{\bf i}-
{\bf a}+{\bf r}_6,\sigma} 
\nonumber\\
&+& y_1 \hat c^{\dagger}_{{\bf i}+{\bf r}_1,\sigma}+
y_2 \hat c^{\dagger}_{{\bf i}+{\bf r}_2,\sigma}+y_3 \hat c^{\dagger}_{{\bf i}+
{\bf r}_3,\sigma}+y_4 \hat c^{\dagger}_{{\bf i}+{\bf r}_4,\sigma}+
y_5 \hat c^{\dagger}_{{\bf i}+{\bf r}_5,\sigma},
\label{E2.14}
\end{eqnarray}
and the equations for the coefficients, based on (\ref{E2.11a}) become
\begin{eqnarray}
&&f_1x_1+f_2x_2=0, \quad a_2x_2+a_4x_4=0, \quad b_2x_2+b_4x_4+b_5x_5=0,
\nonumber\\
&&d_2x_2+d_5x_5+d_6x_6=0, \quad g_6x_6+g_7y_3=0, \quad f_1y_1+f_2y_2=0,
\nonumber\\
&&a_2y_2+a_3y_3+a_4y_4=0, \quad b_2y_2+b_4y_4+b_5y_5=0, \quad
d_2y_2+d_5y_5=0.
\label{E2.15}
\end{eqnarray}
The system (\ref{E2.15}) allows for arbitrary $x_2$ the solution
\begin{eqnarray}
&&x_1=-\frac{f_2}{f_1}x_2, \quad x_4=-\frac{a_2}{a_4}x_2, \quad
x_5=\frac{a_2b_4-b_2a_4}{a_4b_5}x_2, \quad
x_6=-\frac{d_2b_5a_4+d_5(a_2b_4-b_2a_4)}{d_6b_5a_4}x_2,
\label{E2.16}\\
&&y_1=-\frac{Z f_2}{f_1}x_2, \quad y_5=-\frac{Z d_2}{d_5}x_2, \quad
y_4=-\frac{b_2d_5-d_2b_5}{b_4d_5}Z x_2, \quad
y_3=-\frac{a_2b_4d_5-a_4(b_2d_5-d_2b_5)}{a_3b_4d_5}Z x_2,
\nonumber\\
&&y_2=-Z x_2, \quad Z=- \frac{g_6a_3b_4d_5}{g_7d_6a_4b_5} \frac{[
d_2a_4b_5+d_5(a_2b_4-b_2a_4)]}{[a_2b_4d_5+a_4(d_2b_5-b_2d_5)]}=
-\frac{\bar t_ne^{2i\delta'}}{t_ct_ne^{-i\delta_n}} \frac{[
d_2a_4b_5+d_5(a_2b_4-b_2a_4)]}{[a_2b_4d_5+a_4(d_2b_5-b_2d_5)]}.
\nonumber
\end{eqnarray}

\subsection{The deduced ground state}

Based on the considerations from the previous subsection, the 
unnormalized ground state wave vector has the expression
\begin{eqnarray}
|\Psi_g\rangle = \prod_{{\bf i}=1}^{N \leq N_c} \hat B^{\dagger}_{{\bf i},
\sigma_{\bf i}}|0\rangle,
\label{E2.17}
\end{eqnarray}
where the $\hat B^{\dagger}_{{\bf i},\sigma_{\bf i}}$ operators are given in 
(\ref{E2.14},\ref{E2.16}), $N_c$ represents the number of cells in the
system taken with periodic boundary conditions, and $N$ is the total number 
of carriers. 
For $N=N_c$, one has $\sigma_{\bf i}=\sigma$, hence the spin projection being 
fixed, the system is ferromagnetic. This is because neighboring operators are
``connected'', hence the connectivity condition is satisfied, and the unique 
way to obtain $\hat H_{int} |\Psi_g\rangle =0$ is to drop the double occupancy,
hence to have $\sigma_{\bf i}=\sigma$. The obtained ferromagnetic ground state
is localized in the thermodynamic limit.

For $N < N_c$ the ground state is constructed from ferromagnetic clusters whose
cluster moment has arbitrary orientation, hence the ground state is 
paramagnetic. In the thermodynamic limit this state is also localized.

The presented ground states emerge when the conditions (\ref{E2.10}) are
satisfied \cite{Obs1}. 
As (\ref{E2.8}) shows, these conditions are dependent on the
external magnetic field, and through $\epsilon_{\bf j}$, also dependent on the
site selective electric potential. In particular, for zero external fields,
at $t_n < 0$, ferromagnetic solution does not exist \cite{MT3,MT3c}.
Consequently, as (\ref{E2.10}) shows, in the presented case the external 
fields are able to double the parameter space region where the ordered phase 
emerges.

\section{Exact ground states in the interacting case. II.}

As explained in details in Ref. \cite{Obs2}, in a non-integrable case, 
usually is not possible to obtain the desired solution for the ground state
in a single mathematical expression valid for the whole parameter space.
Ground state solutions can be written only in a restricted parameter domain 
which is fixed by the obtained solution of the matching equations. On its turn,
the matching equations are connected to a given decomposition of the Hamiltonian
in positive semidefinite form. Consequently, other decompositions provide
ground state solutions in other regions of the parameter space. 

We explicitly use this observation below by presenting a new decomposition of
the same starting Hamiltonian in positive semidefinite form.

\begin{figure} [h]                                                         
\centerline{\includegraphics[width=6cm,height=5cm]{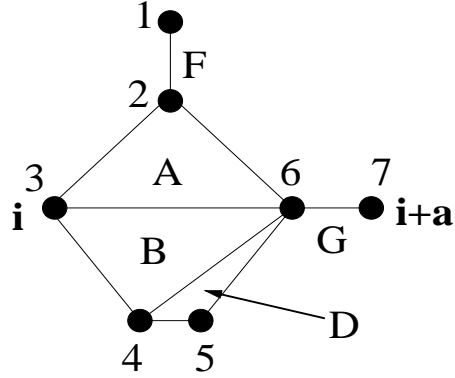}} 
\caption{The five block operators used in the new decomposition. 
Three (A,B,D) are defined on triangles, while two (F,G) on bonds. The
arrow shows the position of the D block.}
\end{figure}

\subsection{The second decomposition step}

We use again 5 block operators for decomposition at $t_p=0$ 
for each cell, but now the blocks are constructed as shown in Fig.5. One 
has in this case
\begin{eqnarray}
&& \hat A_{{\bf i},\sigma}= a_2 \hat c_{{\bf i}+{\bf r}_2,\sigma}+
 a_3 \hat c_{{\bf i}+{\bf r}_3,\sigma}+ a_6 \hat c_{{\bf i}+{\bf r}_6,\sigma},
\quad \hat F_{{\bf i},\sigma}=f_1 \hat c_{{\bf i}+{\bf r}_1,\sigma}+
f_2 \hat c_{{\bf i}+{\bf r}_2,\sigma},
\nonumber\\
&& \hat B_{{\bf i},\sigma}= b_3 \hat c_{{\bf i}+{\bf r}_3,\sigma}+
 b_4 \hat c_{{\bf i}+{\bf r}_4,\sigma}+ b_6 \hat c_{{\bf i}+{\bf r}_6,\sigma},
\quad \hat G_{{\bf i},\sigma}=g_6 \hat c_{{\bf i}+{\bf r}_6,\sigma}+
g_7 \hat c_{{\bf i}+{\bf a},\sigma},
\nonumber\\
&& \hat D_{{\bf i},\sigma}= d_4 \hat c_{{\bf i}+{\bf r}_4,\sigma}+
 d_5 \hat c_{{\bf i}+{\bf r}_5,\sigma}+ d_6 \hat c_{{\bf i}+{\bf r}_6,\sigma}.
\label{E2.1x}
\end{eqnarray}
The transformed Hamiltonian again becomes of the form 
\begin{eqnarray}
\hat H - \hat H_{int}= \sum_{{\bf i},\sigma} [ \hat A^{\dagger}_{{\bf i},\sigma}
\hat A_{{\bf i},\sigma} + \hat B^{\dagger}_{{\bf i},\sigma}
\hat B_{{\bf i},\sigma} + \hat D^{\dagger}_{{\bf i},\sigma}
\hat D_{{\bf i},\sigma} + \hat F^{\dagger}_{{\bf i},\sigma}
\hat F_{{\bf i},\sigma} + \hat G^{\dagger}_{{\bf i},\sigma}
\hat G_{{\bf i},\sigma}] -p\hat N,
\label{E2.2x}
\end{eqnarray}
but now the matching equations are different
(again $\epsilon_5=\epsilon_4, \epsilon_3=\epsilon_6$, $t_p=0$ hold),
namely
\begin{eqnarray}
&&te^{i\delta}=a^*_3a_2=a^*_2a_6, \quad t'e^{i\delta'}=b^*_4b_3=d^*_6d_5,
\nonumber\\
&&t_{6,3}=t_p=a^*_6a_3+b^*_6b_3 = 0, \quad t_{6,4}=b^*_6b_4+d^*_6d_4=0, \quad
t_ne^{i\delta_n}=d^*_5d_4,
\nonumber\\
&&t_f=f^*_2f_1, \quad t_c=g^*_7g_6, \quad \epsilon_1+p=|f_1|^2, \quad
\epsilon_2+p=|f_2|^2+|a_2|^2,
\nonumber\\
&&\epsilon_3+p=|a_3|^2+|g_7|^2+|b_3|^2=|a_6|^2+|b_6|^2+|d_6|^2+|g_6|^2,
\nonumber\\
&&\epsilon_4+p=|b_4|^2+|d_4|^2=|d_5|^2.
\label{E2.3x}
\end{eqnarray}
Solving the matching equations, 11 parameters can be directly and
explicitly expressed
\begin{eqnarray}
&&a_2=\sqrt{\epsilon_2+p-\frac{t_f^2}{\epsilon_1+p}}, \quad
a_3=\frac{te^{-i\delta}}{\sqrt{\epsilon_2+p-\frac{t_f^2}{\epsilon_1+p}}},
\quad a_6=\frac{te^{+i\delta}}{\sqrt{\epsilon_2+p-\frac{t_f^2}{\epsilon_1+p}}},
\nonumber\\
&&b_4=\sqrt{\epsilon_4+p-\frac{t_n^2}{\epsilon_4+p}}, \quad
b_3=\frac{t'e^{+i\delta'}}{\sqrt{\epsilon_4+p-\frac{t_n^2}{\epsilon_4+p}}},
\quad b_6=-\frac{t^2}{t'} \frac{\sqrt{\epsilon_4+p-\frac{t_n^2}{\epsilon_4+p}}}{
\epsilon_2+p-\frac{t_f^2}{\epsilon_1+p}}e^{+i(2\delta+\delta')},
\nonumber\\
&&d_4=\frac{t_ne^{+i\delta_n}}{\sqrt{\epsilon_4+p}}, \quad
d_6=\frac{t'e^{-i\delta'}}{\sqrt{\epsilon_4+p}}, \quad
d_5=\sqrt{\epsilon_4+p},
\quad f_1=\sqrt{\epsilon_1+p}, \quad f_2=\frac{t_f}{\sqrt{\epsilon_1+p}},
\label{E2.4x}
\end{eqnarray}
and one remains with the following four equations written for
the remaining three ($g_6, g_7, p$) unknown variables
\begin{eqnarray}
&&g_7=\sqrt{\epsilon_3+p-|a_3|^2-|b_3|^2}, \quad g_6=\frac{t_c}{g_7},
\nonumber\\ 
&&b_6^*b_4+d_6^*d_4=0,
\quad \epsilon_3+p=|a_6|^2+|b_6|^2+|d_6|^2+|g_6|^2.
\label{E2.5x}
\end{eqnarray}
The first equality of the second line from (\ref{E2.5x}) gives
\begin{eqnarray}
\frac{\epsilon_4+p-\frac{t_n^2}{\epsilon_4+p}}{\epsilon_2+p-\frac{t_f^2}{
\epsilon_1+p}}=\frac{{t'}^2}{t^2}\frac{\bar t_n}{\epsilon_4+p},
\label{E2.6x}
\end{eqnarray}
which, because it is satisfied only at $\bar t_n > 0$, provides back 
the conditions relating the external magnetic field from (\ref{E2.8}). 
Furthermore, (\ref{E2.6x}) determines the
$p$ value as well. The second equality from (\ref{E2.5x}) provides
\begin{eqnarray}
\epsilon_3+p-\frac{t^2}{\epsilon_2+p-\frac{t_f^2}{\epsilon_1+p}}(1+\frac{
\bar t_n}{\epsilon_4+p})=\frac{t_c^2}{\epsilon_3+p-\frac{t^2}{\epsilon_2+p-
\frac{t_f^2}{\epsilon_1+p}}(1+\frac{\epsilon_4+p}{\bar t_n})}+\frac{{t'}^2}{
\epsilon_4+p},
\label{E2.7x}
\end{eqnarray}
which becomes a conditions for the solution of the matching equation.
Together with (\ref{E2.6x},\ref{E2.7x}), the following inequalities must hold
in order to have the presented solution of the matching equations
\begin{eqnarray}
&&\epsilon_1+p>0, \quad (\epsilon_1+p)(\epsilon_2+p)>t_f^2, \quad
\epsilon_4+p>|t_n|, 
\nonumber\\
&&\epsilon_3+p> \frac{t^2(\epsilon_1+p)}{(\epsilon_1+p)(\epsilon_2+p)-t_f^2}
+ \frac{{t'}^2(\epsilon_4+p)}{(\epsilon_4+p)^2-t_n^2}, \quad
\bar t_n > 0.
\label{E2.8x}
\end{eqnarray}
The inequalities (\ref{E2.8x}) define the parameter space region where the
second deduced ground state solution is valid.
 
\subsection{Deduction of the second ground state}

The used procedure is the same as applied in Section IV.

\subsubsection{The possibility of a disconnected solution}

For a possible disconnected solution the $\hat B^{\dagger}_{{\bf i},\sigma}$
operator must have again the expression presented in Eq.(\ref{E2.11}),
hence in the presented case, instead of (\ref{E2.12}) one finds
\begin{eqnarray}
f_1x_1+f_2x_2=0, \quad d_4x_4+d_5x_5=0, \quad x_2a_2=0, \quad x_4 b_4 =0,
\label{E2.9x}
\end{eqnarray}
where now the block operator parameters are taken from (\ref{E2.4x}).
But (\ref{E2.9x}) allows only the solution $x_1=x_2=x_4=x_5=0$, hence 
disconnected solution does not exist.

\subsubsection{The connected solution}

In this case the $\hat B^{\dagger}_{{\bf i},\sigma}$ operator has again the form
presented in (\ref{E2.14}), while now, because the block operators constructing
$\hat H$ are different, instead of (\ref{E2.15}) one obtains
\begin{eqnarray}
&&f_1x_1+f_2x_2=0, \quad a_2x_2+a_6x_6=0, \quad b_6x_6+b_4x_4=0,
\nonumber\\
&&d_4x_4+d_5x_5+d_6x_6=0, \quad g_6x_6+g_7y_3=0, \quad f_1y_1+f_2y_2=0,
\nonumber\\
&&a_2y_2+a_3y_3=0, \quad b_3y_3+b_4y_4=0, \quad d_4y_4+d_5y_5=0,
\label{E2.10x}
\end{eqnarray}
where again the block operator parameters have to be taken from (\ref{E2.4x})
and the first line of (\ref{E2.5x}). The system of equations (\ref{E2.10x}),
for an arbitrary $x_6$ presents the non-trivial solution
\begin{eqnarray}
&&x_1=\frac{a_6f_2}{a_2f_1}x_6, \quad x_2=-\frac{a_6}{a_2}x_6, \quad 
x_4=-\frac{b_6}{b_4}x_6, \quad x_5=\frac{b_6d_4-d_6b_4}{d_5} x_6,
\nonumber\\
&&y_1=-\frac{a_3g_6f_2}{a_2g_7f_1}x_6, \quad y_2=\frac{a_3g_6}{a_2g_7}x_6, \quad
y_3=-\frac{g_6}{g_7} x_6, \quad y_4=\frac{b_3g_6}{b_4g_7}x_6, \quad
y_5=-\frac{b_3g_6d_4}{b_4g_7d_5}x_6.
\label{E2.11x}
\end{eqnarray}
Consequently, in the here presented case one has again a connected solution
of the form (\ref{E2.14}) whose coefficients are provided by (\ref{E2.11x}).

\subsection{The deduced ground state}

The deduced ground state in this Section is again of the form (\ref{E2.17})
but is present on a different region of the parameter space specified by
(\ref{E2.6x},\ref{E2.7x},\ref{E2.8x}). Note that now, as explained above,
the expression of $\hat B^{\dagger}$ operators present in (\ref{E2.17}) is
different, only the qualitative form of the ground state remains the same.
The physical properties of 
$|\Psi_g\rangle$ are similar to the physical properties of the ground state
deduced in the previous Section. In particular, the ordered phase present
in the absence of external fields at $t_n > 0$, is extended also in
the $t_n < 0$ region when external magnetic field is present, which practically
double the parameter space domain of the ferromagnetic phase. This fact
has been demonstrated in two different phase diagram regions in the last two
Sections, hence the extension of the
ordered phase created by the presence of external fields is considerably
higher.

\section{Summary and conclusions}

We investigate pentagon chains described by Hubbard type of Hamiltonians in 
external magnetic and electric fields. The system is in fact a conducting
and organic polymer with pentagon cell not containing magnetic atoms at all.
The external magnetic field perpendicular to the plane of the cell is taken 
into account via Peierls phase factors multiplying the hopping terms of the
Hamiltonian, while the external electric fields by site selective electric
potentials modifying the on-site one-particle potentials. In the presented
conditions one shows that the external fields have a substantial effect on
the physical properties of the chains. First, this modifies considerably
the bare band structure of the system provided by the one-particle part of 
the Hamiltonian, and second, the emergence domains of condensed phases are
also redrawn. We exemplify this by deducing ferromagnetic ground states
in two different regions of the parameter space and show that the parameter 
space domain where the ordered phase occurs is doubled. Since the studied
system is non-integrable and strongly correlated, the deduction process of the
multielectronic ground states is non-approximated, and is based on a technique
which transforms the Hamiltonian in a positive semidefinite form.   

\section{Acknowledgments}

Z. Gul\'acsi kindly acknowledges financial support provided by the 
Alexander von 
Humboldt Foundation, OTKA-K-100288 (Hungarian Research Funds for Basic 
Research) and TAMOP 4.2.2/A-11/1/KONV-2012-0036 (co-financed by EU and European 
Social Fund). 

\appendix

\section{The calculation of the Peierls phase factors}

The calculation of the Peierls phase factors is performed by using 
Eq.(\ref{UE1}). For the vector potential one uses the Landau gauge expression,
namely ${\bf A}=(A_x=0, A_y=x B, A_z=0)$, where $B=|{\bf B}|$, and the used
system of coordinates $(x0y)$ is presented in Fig.1c. Note that 
${\bf B}=constant$ holds, and ${\bf B}$ is directed along the z-axis 
perpendicular to the (x0y) plane in Fig.1c.

The calculations provide the following Peierls phase factors:

1) Along the bond $(1 \to 2)$, because the vectors ${\bf A}$ and $d{\bf l}$ are
orthogonal, one has ${\bf A} \cdot d{\bf l}=0$, hence $\phi_{1,2}=0$ holds.

2) Along the bonds $(3 \to 6)$, and  $(6 \to 7)$, because of $x=0$, one has
${\bf A}=0$, hence $\phi_{3,6}=\phi_{6,7}=0$.

3) Along the bond  $(4 \to 5)$, $A_y= a_2 B$=constant, consequently
\begin{eqnarray}
\int_{i_4}^{i_5} {\bf A} d{\bf l}=a_2 B \int_{y_4}^{y_5} dy =a_2 b_1 B.
\label{UE4}
\end{eqnarray}
Please note that $y_5-y_4=b_1$, see Fig.1a,
consequently $\phi_{4,5}=\frac{2\pi}{\Phi_0} a_2 b_1 B =\delta_n$.

4) Along the bond $(3 \to 4)$, denoting by $\alpha$ the angle between the
bond $(3 \to 4)$ and $0x$ axis [$\cos \alpha =x_4/L, \sin \alpha =y_4/L$,
where $L=\sqrt{x_4^2+y_4^2}$ is the lenght of the bond $(3 \to 4)$], taking
$A_{3,4}=A\cos(\frac{\pi}{2}-\alpha)=A \sin \alpha$, $A=|{\bf A}|$, as the
projection of the vector potential along the bond $(3 \to 4)$, and denoting
by $l_{3,4}$ the ${\bf l}$ variable along the bond $(3 \to 4)$, one obtains
\begin{eqnarray}
\int_{i_3}^{i_4} {\bf A} d{\bf l} &=& \int_0^L A_{3,4} d l_{3,4} =
\int_0^L A \sin \alpha d l_{3,4} = B \sin \alpha \cos \alpha \int_0^L l_{3,4}
d l_{3,4} = B \sin \alpha \cos \alpha \: \frac{L^2}{2}
\nonumber\\
&=& \frac{B}{2} x_4 y_4=
\frac{a_2 b_2 B}{2},
\label{UE5}
\end{eqnarray}
where one had $x_{3,4}=l_{3,4} \cos \alpha$, $A=x_{3,4}B=l_{3,4} B \cos \alpha$,
and $x_4=a_2, y_4=b_2$, see Fig.1a.
Consequently, one obtains $\phi_{3,4}=\frac{2\pi}{\Phi_0} \frac{a_2 b_2 B}{2} 
=\delta'$.

5) The same result is obtained along the bond $(5 \to 6)$. Indeed, if $l_{6,5}$,
the ${\bf l}$ variable along the bond $(6 \to 5)$, is measured from the site 6,
taking into account that now ${\bf A} d{\bf l} < 0$, one has
\begin{eqnarray}
\int_{i_6}^{i_5} {\bf A} d{\bf l} &=& -\int_0^L A_{6,5} d l_{6,5} =-
\int_0^L A \sin \alpha d l_{6,5} = - B \sin \alpha \cos \alpha \int_0^L l_{6,5}
d l_{6,5} = - B \sin \alpha \cos \alpha \: \frac{L^2}{2}
\nonumber\\
&=& -\frac{a_2 b_2 B}{2}.
\label{UE6}
\end{eqnarray}
This means $\phi_{6,5}=-\delta'$, consequently $\phi_{5,6}=-\phi_{6,5}=\delta'$.

6) Along the bond $(3 \to 2)$, denoting by $\beta$ the angle between the 
bond $(3 \to 2)$ and 0x axis $[ \cos \beta =|x_2|/L', \sin \beta = y_2/L',
L'=\sqrt{x_2^2+y_2^2} ]$, taking the ${\bf l}$ variable along the bond 
$(3 \to 2)$ to be $l_{3,2}$ as measured from the origin (site 3), observing
that the ${\bf A}$ projection to the bond $(3 \to 2)$ is 
$A_{3,2}=A \cos(\frac{\pi}{2}-\beta)=A \sin \beta$, but now 
$x_{3,2}=-l_{3,2} \cos \beta$ because we are placed on the negative region of the
0x axis, hence $A=B x_{3,2}=-B l_{3,2} \cos \beta$, one has
\begin{eqnarray}
\int_{i_3}^{i_2} {\bf A} d{\bf l} &=& \int_0^{L'} A_{3,2} d l_{3,2} =
\int_0^{L'} A \sin \beta d l_{3,2} = - B \sin \beta \cos \beta \int_0^{L'} l_{3,2}
d l_{3,2} = - B \sin \beta \cos \beta \: \frac{{L'}^2}{2}
\nonumber\\
&=& -\frac{|x_2| y_2 B}{2}.
\label{UE7}
\end{eqnarray}
Consequently, taking into account that $|x_2|=a_1, y_2=b/2$ where $b=2b_2+b_1$,
one obtains $\phi_{3,2}=-\frac{2\pi}{\Phi_0} \frac{a_1 b B}{4} =-\delta$.
But since $\phi_{2,3}=-\phi_{3,2}$, we have as result
$\phi_{2,3}=\frac{2\pi}{\Phi_0} \frac{a_1 b B}{4} =\delta$.

7) Along the bond $(6 \to 2)$ we obtain the result given in $\phi_{2,3}$.
This is because now the projection of ${\bf A}$ on the bond with the $l_{6,2}$
orientation is negative, i.e. $A_{6,2}=-A \sin \beta$, but $x_{6,2}$ remains
$x_{6,2}=-l_{6,2} \cos \beta$. Consequently
\begin{eqnarray}
\int_{i_6}^{i_2} {\bf A} d{\bf l} &=& \int_0^{L'} A_{6,2} d l_{6,2} =-
\int_0^{L'} A \sin \beta d l_{6,2} = + B \sin \beta \cos \beta \int_0^{L'} l_{6,2}
d l_{6,2} = B \sin \beta \cos \beta \: \frac{{L'}^2}{2}
\nonumber\\
&=& \frac{|x_2| y_2 B}{2}.
\label{UE8}
\end{eqnarray}
This means that $\phi_{6,2}=\frac{2\pi}{\Phi_0} \frac{a_1 b B}{4} =\phi_{2,3}=
\delta$.

8) Along the bond  $(2 \to 5)$ the Peierls factor is obtained as follows.
Let introduce the $l$ axis connecting the sites 2 and 5. This axis intersects
the translated x axis as shown in Fig. 6. Let us denote by $\gamma$ the angle 
between the x axis and $l$ axis.

\begin{figure} [h]                                                         
\centerline{\includegraphics[width=4cm,height=4cm]{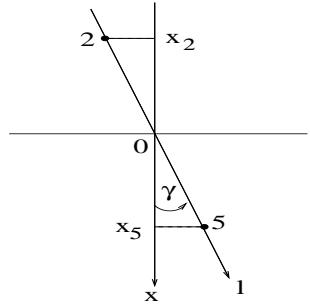}} 
\caption{The $l$ axis along the line connecting the sites 2 and 5. This
axis is rotated by the angle $\gamma$ relative to the x axis.}      
\end{figure}                                                               

One has $|x_2|=a_1, x_5=a_2, \cos \gamma=(a_1+a_2)/L_{2,5}, \sin \gamma= 
b_1/(2 L_{2,5}$ (see Fig.1a), where the distance between the sites 2 and 5 is
$L_{2,5}=\sqrt{(a_1+a_2)^2+(b_1/2)^2}$. The connection between the coordinate x
and the coordinate $l$ is $x=l \cos \gamma$, hence $l_2=x_2/\cos \gamma,
l_5= x_5/\cos \gamma$ holds. The projection of the vector potential on the
$l$ axis is $A_l=A \sin \gamma$, and because $A=x B = ( l \cos \gamma) B$,
one has $A_l=B l \cos \gamma \sin \gamma$. Consequently
\begin{eqnarray}
\int_{i_2}^{i_5} {\bf A} d{\bf l} &=& \int_{l_2}^{l_5} A_l d l =
B (\sin \gamma \cos \gamma) \int_{l_2}^{l_5} l dl=
B (\sin \gamma \cos \gamma)  \frac{{l_5}^2-{l_2}^2}{2}
\nonumber\\
&=& \frac{B}{2} (\frac{{x_5}^2}{\cos^2 \gamma} -\frac{{x_2}^2}{\cos^2 \gamma}) 
 \sin \gamma \cos \gamma = \frac{B}{2} (x_5^2-x_2^2)\frac{\sin \gamma}{
\cos \gamma}
\nonumber\\
&=& \frac{B}{2} (a_2^2-a_1^2) \frac{b_1}{2(a_1+a_2)}=
B \frac{(a_2-a_1)b_1}{4}.
\label{UE9}
\end{eqnarray}
Consequently $\phi_{2,5}=\frac{2\pi}{\Phi_0} \frac{(a_2-a_1)b_1}{4} B =
\delta_1$.

9) Along the bond  $(4 \to 2)$ one obtains the same result. Indeed, now
the integrals must be taken from $l_5=l_4$ to $l_2$, but $l=-x \cos \gamma$
since the $l$ axis is directed from site 4 to the site 2, hence the two 
negative signs compensate each other. Consequently $\phi_{4,2}=\delta_1$. 

Summarizing the obtained results one has
\begin{eqnarray}
\delta &=& \frac{2\pi}{\Phi_0}B \frac{a_1b}{4}=\phi_{6,2}=\phi_{2,3}, \quad
\delta'= \frac{2\pi}{\Phi_0}B \frac{a_2b_2}{2}=\phi_{3,4}=\phi_{5,6},
\nonumber\\
\delta_1 &=& \frac{2\pi}{\Phi_0}B \frac{b_1(a_2-a_1)}{4}=\phi_{4,2}=\phi_{2,5},
\quad \delta_n =\frac{2\pi}{\Phi_0}B a_2 b_1=\phi_{4,5},
\nonumber\\
0 &=& \phi_{3,6}=\phi_{6,7}=\phi_{1,2}.
\label{UE10x}
\end{eqnarray}

Now one check the calculations based on (\ref{UE3}).
The total surface of the pentagon $S=S_{3,6,2}+S_{3,4,5,6}$ 
is composed from the surface of the 
upper triangle $S_{3,6,2}=a_1b/2=a_1(2b_2+b_1)/2$, and the surface of the lower
trapezoid $S_{3,4,5,6}=a_2(b+b_1)/2=a_2(2b_2+2b_1)/2$, consequently, one finds
$S=(1/2)[a_1(2b_2+b_1)+a_2(2b_2+2b_1)]$. Furthermore the middle triangle 
$(4,5,2)$ enclosed into the pentagon has the surface $S_{4,5,2}=b_1(a_1+a_2)/2$,
hence the two lateral triangles with the same area $(3,4,2)$ and $(2,5,6)$ have
the surface $S_{3,4,2}=S_{2,5,6}=(S-S_{4,5,2})/2=[b_1a_2+2b_2(a_1+a_2)]/4$. Now one 
can check that as required by Eq.(\ref{UE3}) one has

\begin{eqnarray}
&&\phi_{2,3}+\phi_{3,4}+\phi_{4,5}+\phi_{5,6}+\phi_{6,2}=\frac{2\pi}{\Phi_0}B S,
\nonumber\\
&&\phi_{2,3}+\phi_{3,6}+\phi_{6,2}=\frac{2\pi}{\Phi_0}B S_{3,6,2},
\nonumber\\
&&\phi_{3,4}+\phi_{4,5}+\phi_{5,6}+\phi_{6,3}=\frac{2\pi}{\Phi_0}B S_{3,4,5,6},
\nonumber\\
&&\phi_{2,5}+\phi_{5,6}+\phi_{6,2}=\frac{2\pi}{\Phi_0}B S_{2,5,6},
\nonumber\\
&&\phi_{2,3}+\phi_{3,4}+\phi_{4,2}=\frac{2\pi}{\Phi_0}B S_{3,4,2},
\nonumber\\
&&\phi_{4,5}-\phi_{2,5}-\phi_{4,2}=\phi_{4,5}+\phi_{5,2}+\phi_{2,4}=
\frac{2\pi}{\Phi_0}B S_{4,5,2}.
\label{UE11}
\end{eqnarray}
Consequently, the calculated Peierls phase factors are correct.

\section{The $|a_3|=|d_6|$ condition}

From the subsection B.1 it seems at first view that the condition $|a_3|=|d_6|$
is artificially imposed, so provides only a particular solution of
(\ref{E2.5}). This impression is not true. This fact will be demonstrated
below. For this reason one concentrates below on the last equation from
(\ref{E2.5}), namely that written for $\epsilon_4+p$. Indeed, without 
presumptions, from (\ref{E2.6}) one has 
\begin{eqnarray}
|b_4|^2=\bar t_n \frac{|d_6|^2}{|a_3|^2}. 
\label{E2.18}
\end{eqnarray}
This
introduced in the first equality of $\epsilon_4+p$ gives
\begin{eqnarray}
|a_3|^2=\frac{\bar t_n|d_6|^2+{t'}^2}{\epsilon_4+p}.
\label{E2.19}
\end{eqnarray}.
But the same (\ref{E2.18}) introduced in the second equality of $\epsilon_4+p$
gives
\begin{eqnarray}
|a_3|^2=\frac{\bar t_n}{t_n^2}[(\epsilon_4+p)-{t'}^2].
\label{E2.20}
\end{eqnarray}
Now taking $|a_3|^2=|a_3|^2$ from (\ref{E2.19}) and (\ref{E2.20}) one obtains
(note that ${\bar t_n}^2=t_n^2$ holds):
\begin{eqnarray}
|d_6|=\frac{|t'|}{\sqrt{\epsilon_4+p-\bar t_n}}.
\label{E2.21}
\end{eqnarray}
This relation introduced in (\ref{E2.19}) provides
\begin{eqnarray}
|a_3|=\frac{|t'|}{\sqrt{\epsilon_4+p-\bar t_n}}.
\label{E2.22}
\end{eqnarray}
As seem, $|a_3|=|d_6|=\frac{|t'|}{\sqrt{\epsilon_4+p-\bar t_n}}$
is not a simplification of the solution, is the only existing solution of
(\ref{E2.5}).



\begin{thebibliography}{00}

\bibitem{App0}
J. D. Stenger-Smith, Prog. Polym. Sci. {\bf 23}, 57 (1998).

\bibitem{App1}
A. Ramanaviciusa, A. Ramanavicienea and A. Malinauskasc,
Electrochimica Acta {\bf 51}, 6025 (2006).

\bibitem{App2}
T. Liu, L. Finn, M. Yu, H. Wang, T. Zhai, X. Lu, Y. Tong and Y. Li, 
Nano Lett. {\bf 14}, 2522 (2014).

\bibitem{App3}
D. D Ateh, H.A. Navsarin and P. Vadgama, Interface {\bf 3}, 741 (2006).

\bibitem{App4}
D. Martindale, New Sci. {\bf 182}, 38 (2004).

\bibitem{App5}
P. Saville, Defence R\& D Canada, DRDC Atlantic TM 2005-004, (2005)

\bibitem{App6}
C. Wang, W. Zheng, Z. Yue, C. O. Too and G. G. Wallace, Advanced Materials  23, 
3580, (2011).


\bibitem{App7}
Z. Wang, P. Tammela, P. Zhang, J. Huo, F. Ericson, M. Stromme and L. Nyholm,  
Nanoscale, 2014,6, 13068-13075

\bibitem{App8}
S. Hara, T. Zama, W. Takashima and K. Kaneto,
Polymer Journal (2004) 36, 151-161;


\bibitem{App9}
S. P. Lim, A. Pandikumar, Y. S. Lim, N. M. Huang and H. N. Lim,
Scientific Reports {\bf 4}, 5305 (2014).

\bibitem{App10}
Q. Wang, J. Wang, G. Lv, F. Wang, X. Zhou, J. Hu and Q. Wang,
Journal of Materials Science {\bf 49}, 3484 (2014).


\bibitem{App11}
D. Samanta, J. L. Meiser and R. N. Zare,
Nanoscale. {\bf 7}, 9497-504 (2015). 

\bibitem{App12}
A. Nan, I. Craciunescu and R. Turcu, in
{\it  Aspects on Fundaments and Applications of Conducting Polymers}, 
pg. 159-182, Chapt. 8: Conducting Polypyrrole Shell as a Promis
ing Covering for Magnetic Nanoparticles,
 pg. 159-182, Chapt. 8: INTECH publication, Rijeka- Shanghai, 2010.


\bibitem{App13}
Y. Pang, N. Chen, and L. Hong, Mat. Res. Soc. Symp. Proc. 
{\bf 635}, pg. C69.1, (2001).

\bibitem{Intr8}
Z. Gul\'acsi, Int. Jour. Mod. Phys. B. {\bf 27}, 1330009 (2013).


\bibitem{Intr25}
Z. Gul\'acsi, A. Kampf, D. Vollhardt, Phys. Rev. Lett. {\bf 105}, 266403 (2010).

\bibitem{Intr25a}
Z. Gul\'acsi, Eur. Phys. Jour. B. {\bf 87}, 143 (2014).


\bibitem{MT3}
M. Gul\'acsi, G. Kov\'acs, and Z. Gul\'acsi, Phil. Mag. Lett. {\bf 94}, 269
(2014).

\bibitem{MT3a}
 M. Gul\'acsi, Gy. Kov\'acs, Zs. Gul\'acsi, Europhysics Lett.
{\bf 107}, 57005, (2014). 

\bibitem{MT3b}
M. Gul\'acsi, Gy. Kov\'acs, Zs. Gul\'acsi, Mod. Phys. Lett. B. 
{\bf 28}, 1450220 (2014).

\bibitem{MT3c}
Z. Gu\'acsi, M. Gul\'acsi, Aditi Jour. of Math. Phys. {\bf 4}, 44 (2013).

\bibitem{STR1}
G. Brocks, J. van den Brink and A. F. Morpurgo, Phys. Rev. Lett. {\bf 93},
146405 (2004)

\bibitem{STR2}
T. O. Wehling et al., Phys. Rev. Lett. {\bf 106}, 236805 (2011).

\bibitem{STR3}
Z. Gul\'acsi, Jour. of Phys. Conf. Ser.  {\bf 410}, 012011 (2013).

\bibitem{PAM1a} 
I. Orlik, Z. Gul\'acsi, Phil. Mag. Lett. {\bf 78}, 177 (1998).

\bibitem{PAM1b}
Z. Gul\'acsi, I. Orlik, Jour. of  Phys. A. {\bf 34}, L359 (2001).

\bibitem{PAM2a}
Z. Gul\'acsi, Phys. Rev. B. {\bf 66}, 165109 (2002);
Eur. Phys. Jour. B {\bf 30}, 295 (2002).

\bibitem{PAM2b}
P. Gurin and Z. Gul\'acsi, Phys. Rev. B. {\bf 64}, 045118 (2001).

\bibitem{PAM3a}
Z. Gul\'acsi and D. Vollhardt, Phys. Rev. Lett. {\bf 91}, 186401 (2003).

\bibitem{PAM3b}
Z. Gul\'acsi and D. Vollhardt, Phys. Rev. B. {\bf 72}, 075130 (2005).

\bibitem{Q1}
Z. Gul\'acsi, A. Kampf, D. Vollhardt, Phys. Rev. Lett. {\bf 99}, 026404 (2007); 
Progr. Theor. Phys. Suppl. {\bf 176}, 1 (2008).

\bibitem{D1}
Z. Gul\'acsi, M. Gul\'acsi, Phys. Rev. B. {\bf 73}, 014524 (2006).

\bibitem{D2}
Z. Gul\'acsi, Phys. Rev. B. {\bf 77}, 245113 (2008).

\bibitem{D3}
Z. Gul\'acsi, Phys. Rev. B. {\bf 69}, 054204 (2004).

\bibitem{Fet1}
Jong-in Hahm, Sensors (Basel) {\bf 11}, 3327 (2011). 

\bibitem{Fet2}
F. Patolsky, G. Zheng, C. M. Lieber, Anal. Chem. {\bf 78}, 4260 (2006).

\bibitem{Fet3}
J. C. Bradley1, H. M. Chen, J. Crawford, J. Eckert, K. Ernazarova, 
T. Kurzeja, M. Lin, M. McGee, W. Nadler and S. G. Stephens,
Nature {\bf 389}, 268 (1997).

\bibitem{Fet4}
M. Cochet, W. K. Maser, A. M. Benito, M. A. Callejas, M. T. Martinez,   
J. M. Benoit, J. Schreiber and  O. Chauvet,  
Chem. Commun. {\bf 16}, 1450 (2001). 

\bibitem{Fet5}
A. P. Esser-Kahn, V. Trang and M. B. Francis,
Jour. Am. Chem. Soc. {\bf 132}, 13264-9 (2010).

\bibitem{Fet6}
B. McCarthy, J. N. Coleman, S. A. Curran, A. B. Dalton, A. P. Davey,
Z. Konya, A. Fonseca, J. B. Nagy and W. J. Blau, Jour. of Mater. Sci. Lett.
{\bf 19}, 2239 (2000).

\bibitem{Fet7}
N. Shida, Y. Koizumi, H. Nishiyama, I. Tomita and S. Inagi, 
Angew. Chem. Int. Ed. {\bf 54}, 3922 (2015). 

\bibitem{Fet8}
Yoshiki Chujo in {\em Macromolecular Design of Polymeric Materials},
Eds. K. Hatada, T. Kitayama and O. Vogl, Marcel Dekker Inc. USA 1997,
Chapt. 18: pg.323-337: Design of Reactive Polymers and their Applications, 
pg. 333.

\bibitem{Rich1}
O. Derzhko, J. Richter, A. Honecker, and R. Moessner, Phys. Rev. B. {\bf 81}, 
014421 (2010).

\bibitem{Rich2}
O. Derzhko and J. Richter, Phys. Rev. B. {\bf 90}, 045152, (2014);

\bibitem{Rich3}
O. Derzhko, J. Richter and M. Maksymenko,
Int. Jour. Mod. Phys. B. {\bf 29}, 1530007 (2015). 

\bibitem{LCR1}
T. Taychatanapat, K. Watanabe, T. Taniguchi and P. Jarillo-Herrero,
Nature Phys. {\bf 7}, 621 (2011).

\bibitem{LCR2}
Z. Dutton, K.V.R.M. Murali, W. D. Oliver and  T.P. Orlando,
Phys. Rev. B. {\bf 73}, 104516 (2006) 

\bibitem{LCR3}
A. J. Fendrik, M. J. Sanchez,  arXiv:cond-mat/9811272

\bibitem{LCR4}
G. Sun, X. Wen, M. Gong, D. W. Zhang, Y. Yu, S. L. Zhu, J. Chen, 
P. Wu and S. Han, Scientific Reports {\bf 5}, 8463 (2015). 

\bibitem{LCR5}
M. Assili, S. Haddad, W. Kang, Phys. Rev. B. {\bf 91}, 115422 (2015).

\bibitem{LCR6} 
D. Jacob, Jour. Phys.: Condens. Matter {\bf 27}, 245606 (2015). 

\bibitem{LCR7}
K. D. Belashchenko, J. Weerasinghe, Sai Mu and B. S. Pujari,
Phys. Rev. B. {\bf 91}, 180408 (2015). 

\bibitem{GX1}
Z. Gul\'acsi, Jour. of Phys. Conf. Ser.  {\bf 410}, 012011 (2013).

\bibitem{Obs1}
These conditions represent requirements which provide lower dispersionless
band created by external fields.

\bibitem{Obs2}
R. Trencs\'enyi and Z. Gul\'acsi, 
Phil. Mag.  {\bf 92}, 4657 (2012).
 
\end{thebibliography}

\end{document}